# Self-Retracting Motion of Graphite Microflakes

Quanshui ZHENG[1*], Bo JIANG[1], Shoupeng LIU[2], Jing ZHU[3], Qing JIANG[4], Yuxiang WENG[2], Li LU[2], Sheng WANG[5], Qikun XUE[2], Lianmao PENG[5]

[1] Department of Engineering Mechanics, Tsinghua University, Beijing 100084, China

[2] Institute of Physics, Chinese Academy of Science, Beijing 100080, China

[3] Department of Materials Science and Engineering, Tsinghua University, Beijing 100084, China

[4] Department of Mechanical Engineering, University of California, Riverside, California 92521-0425, USA

[5] Department of Electronics, Peking University, Beijing 100871, P.R. China

* email: zhengqs@tsinghua.edu.cn

**We report the observation of a novel phenomenon, the self-retracting motion of graphite, in which tiny flakes of graphite, after being displaced to various suspended positions from islands of highly orientated pyrolytic graphite, retract back onto the islands under no external influences. Our repeated probing and observing such flakes of various sizes indicate the existence of a critical size of flakes, approximately 3~5 μm, above which the self-retracting motion does not occur under the operation. This helps to explain the fact that the self-retracting motion of graphite has not been reported, because samples of natural graphite are typical larger than this critical size. In fact, reports of this phenomenon have not been found in the literature for single crystals of any kinds. A model that includes the static and dynamic shear strengths, the van der Waals interaction force, and the edge dangling bond interaction effect, was used to explain the observed phenomenon. These findings may conduce to create nano-electromechanical systems with a wide range of mechanical operating frequency from mega to giga hertzs.**

Graphite is one of the most useful materials because of its many extreme mechanical, electrical and thermal properties as well as biocompatibility. For example, due to the superlubricity [1,2] between graphite layers and the extreme high elastic moduli and strengths within the layers, graphite has widely been used as durable solid lubricants. New surprising properties of graphite have been discovered at times, such as the existence of graphite monolayer in the free state [3] and extreme anisotropy owned by graphite compared with all other hexagonal crystalline materials [4]. The former provides an ample scope for fundamental research and new technologies [5] and has already prompted intensive studies, such as designable electrical properties [6-8] and quantum Hall effect [9]. Unlike the carbon nanotubes or other low-dimensional nanostructured materials, graphene nanoribbons with intricate sub-micrometer structures can now be fabricated [10,11], leading to the use of the graphite sheet structure to fabricate electromechanical resonator [12].

The recent experiments on controlled sliding and extraction-releasing of nested shells in individual multiwalled carbon nanotubes (MWNTs) [13,14] revealed that the MWNTs have similar superlubricity as graphite, with the interwall shear strength against sliding ranging from 0.08 to 0.3 MPa. For comparison, the interlayer shear strength values of high quality crystalline graphite range from 0.25 to 0.75 MPa varying with shear directions, and those between rotated graphite layers are one order lower in magnitude [2]. More interestingly, some extracted inner shells were found to self-retract back into the outer shells [14]. Inspired by these observations, MWNT-based oscillators as the first sample nanoelectromechanical system (NEMS) with frequencies in the gigahertz range were proposed [15] and then have been intensively studied [16].

Graphite was named in 1789 from the Greek γραφειν (graphein): "to draw/write", for its use in pencils. When people drawn or wrote using pencils, they produced countless tiny pieces of graphite, each consisting of many graphene sheets [Fig. 1(a)]. Considering that the interlayer interaction of graphite is of the same nature as the interwall interaction of MWNTs, we have wondered whether or not one can generate a similar self-retracting motion for graphite, in view of the fact that its interlayer slipping phenomenon has been known for a long time. We have thus explored this issue, and with this letter, we have documented our detailed observations and analysis.

The experiments were carried out on square graphite-SiO$_2$ islands of the height about 200 nm and various side lengths (sizes), *L*, ranging from 0.5 to 5 μm [Fig. 1(b,c)]. The technique we used to prepare such islands is similar to that reported in [17], as detailed below. As illustrated in Fig. 1(b), we firstly deposited a $SiO_2$ film about 100 nm in thickness by controlling the deposition time onto a freshly cleaved surface (5×5 mm) of a highly orientated pyrolytic graphite (HOPG) sample, using plasma enhanced chemical vapor deposition (PE-CVD). Secondly, we patterned it into squares with negative photoresist (PMMA495), by spin-coating onto the surface of the $SiO_2$ film, and by electron beam lithography. Thirdly, the portion of the $SiO_2$ film, unprotected by the photoresist, was etched away by reactive ion etching (RIE). Fourthly, the remaining portion of the $SiO_2$ film was then used as the masking surface in a follow-up oxygen plasma etching to remove the photoresist. This led to the final graphite-$SiO_2$ square islands, as illustrated in Fig. 1(b).

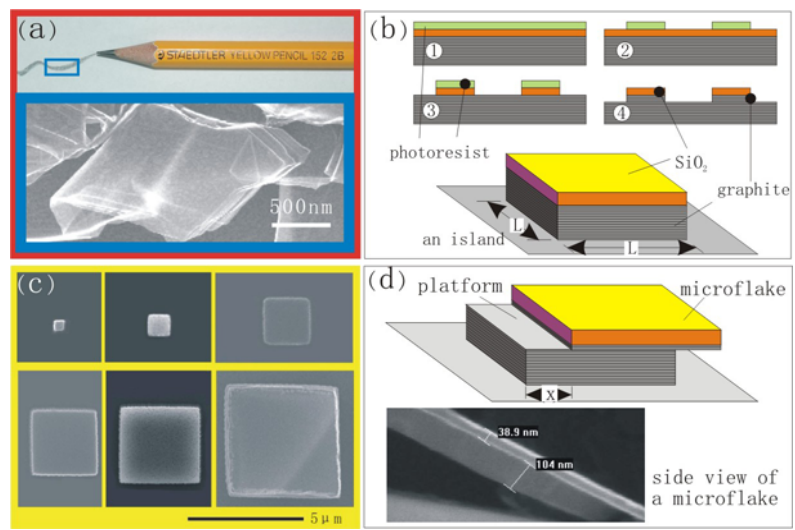

FIG.1 (color online). Microflakes: (a) countless tiny graphite pieces produced as drawing with a 2B pencil. (b) illustrative steps for carving square graphite-$SiO_2$ islands from a highly orientated pyrolytic graphite (HOPG) coated with a $SiO_2$ film. (c) top views of some samples of square islands. (d) an illustrative of a slipped microflake, its graphite platform, and a SEM side view of an overthrew microflake showing it to be consist of a $SiO_2$ film of thickness 104 nm and graphite laminate of thickness 38.8 nm.

The experiment on each tested graphite-$SiO_2$ island was carried out by controlling the probe of a micromanipulator MM3A (Kleindiek) set in a scanning electron microscope (SEM) to laterally push an upper edge or horizontally rub on the top surface of the island [Fig. 2(a,b)]. The probes were selected to have similar tip sizes as the tested islands. Their lateral motions, up to 5 nm in accuracy, were manually controlled by rotating a knob of the micromanipulator. The moving probe and microflakes were monitored in both image and digital. With this method we successfully slipped out a microflake from each of the tested dozens islands of 1 or 2 μm side length to various prescribed suspended positions. From an overthrew 2 μm microflake [Fig. 1(d)] we found that it consists of a 104 nm-thick $SiO_2$ film and a 38.8 nm-thick graphite lamina. Thus, the underlying square platform is purely graphitic and the both contact surfaces are graphite basal planes. Furthermore, we found that each suspended microflake under test can automatically and fully retract back onto the graphite platform top immediately after the applied force is released by removing the probe away from the microflake. The observation that the retracting motion occurred even though the probe was removed away in the direction opposing to the retraction motion direction [Fig. 2(c) and SI-Movie] excludes the qualm that the retracting motion would be caused by the adhesion or electrostatic force between the microflakes and the removing probe, and this thus validates the term "self-retracting motion".

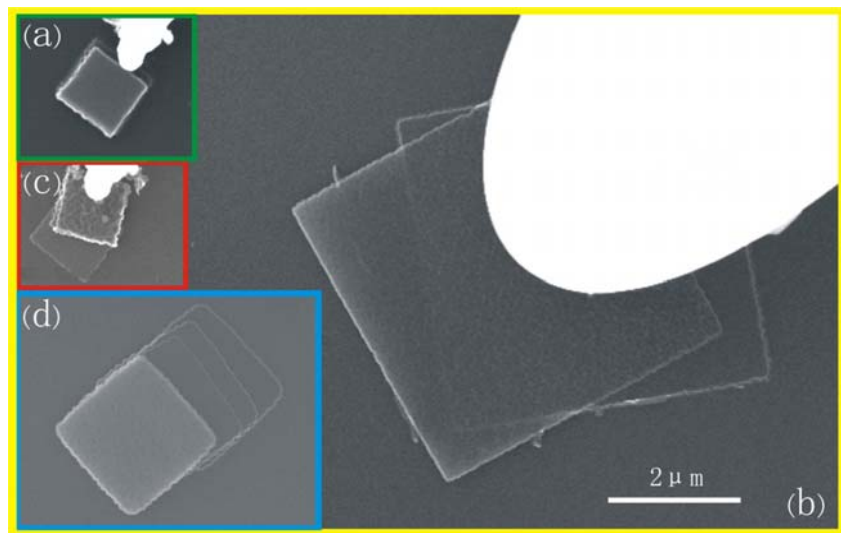

FIG.2 (color online). The slip and self-retraction of microflakes. (a) Slip without rotation. (b) slip with rotation. (c) backward slip. (d) multiflake slip.

The slip and self-retraction processes were easily repeated again and again, and many of them were accompanied with rotations [Fig. 2(b,c)], particularly when the slip forces were eccentrically applied. In addition, the different deformations of the probe, measured by the observed microflake displacements and recorded knob rotations, reveal that the resistance against initiating the slip motion is significantly larger than the resistance during motion. After having either rotational or repeatedly translational slips, the sliding resistance was found to be even smaller. However, in these case that we released the applied force after we had taken a pause of 15~20 seconds to record a high resolution SEM image of a microflake in its suspended position, we found that the microflake stayed at the suspended position and we observed no self-retraction. Interestingly, self-retraction of this microflake was again observed after slipping it further outwards and then immediately releasing it, while the self-retraction motion, however, returned it only to the previous stop position. For a possible explanation, we have noted previous reports [18] that amorphous carbon layers of several nanometers in thickness were formed within 20 seconds on surfaces exposed in electron beams of the SEM. We have also found that repeating the slip-retraction process may generate several graphite microflakes under the top graphite-$SiO_2$ microflake [Fig. 2(d)], and correspondingly, the self-retracting motion become multi-body motion.

It was observed that slipped microflakes are not always self-retracting, even if released immediately after slipping, depending upon the island size and whether or not a microflake was rotated when it is slipped. For islands of 3.5 μm size length, this self-retracting motion was observed for most of the slipped and rotated microflakes, but only for some of the microflakes that were slipped without rotation. For the 5 μm islands, the self-retracting motion was only occasionally observed for slipped and rotated microflakes, but was never observed for the non-rotated ones. This indicates that the probability of the self-retraction decreases substantially with the increasing side length of the microflakes. We have tested 15 slipped microflakes for each island size, and our observations indicates that probabilities were 100% for 1 or 2 μm islands, and 87%, 33%, and 13%, respectively, for 3.0, 3.5 and 5.0 μm islands. These observations suggest the existence of rotation-dependent critical or maximum sizes under the described slip operation, approximately 3~5 μm, to permit the self-retracting motion. This helps to explain the fact that the graphite self-retracting motion was not previously observed because samples of natural graphite are typically larger than the critical sizes. In addition, our attempts of slipping out microflakes from 0.5 μm size islands, however, had always led to overthrow the islands. Furthermore, we found that the stiff SiO$_2$ coats play a key role for realizing the self-retracting motion. After having removed the SiO$_2$ coats from some islands of various side lengths, we failed to slide out any microflakes without rolling or buckling the graphite monolayers on the coat-removed islands [Fig. 3(a)].

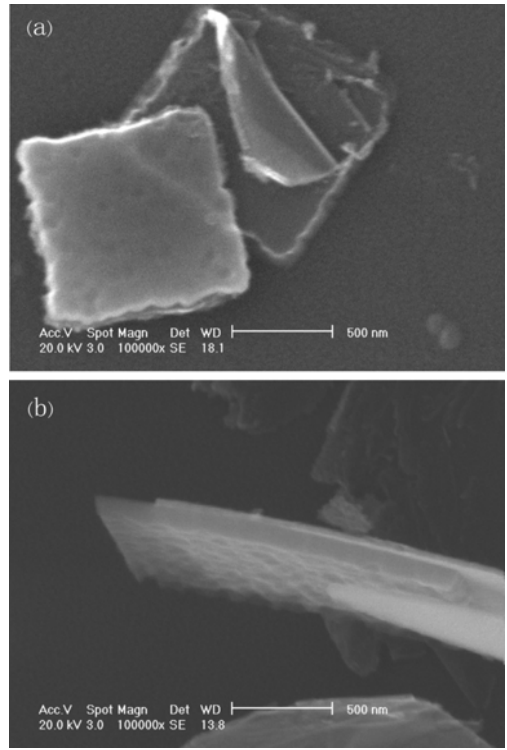

FIG.3. Failed microflakes. (a) Slipping on SiO$_2$ cover-removed islands leaded to rolling and other damages. (b) The wrapped microflake slipped from a 5 mm island and did not self-retract.

To further explain these observations, we first consider a simple self-retracting motion of a microflake that was slipped out from an island of side length $L$ along a side direction without rotation [Fig. 1(d)]. The forces affecting the retracting motion are analyzed below. Slipping out the microflake to a distance $x$ creates two graphite basal surfaces with the total area $2Lx$. Thus, the microflake-platform system has an excess potential energy, $2Lx\gamma$, with the graphite basal surface energy $\gamma$ measured to be 0.12 Jm$^{-2}$ [19], and consequently the suspended microflake is pulled backward its non-slip position by the platform with a constant force (solid-solid "capillary" force), $F_d = 2L\gamma$. The retraction driving force $F_d$ may be further enhanced by the interaction between the edge dangling bonds and graphite basal planes and the total retraction driving force may be modeled [20] as $F_d^* = 2L\gamma^*$ with $\gamma^* = \gamma + \gamma_e$, where $F_e = 2L\gamma_e$ was called as the edge effect force and $\gamma_e$ was measured to be 0.45 and 0.67 Jm$^{-2}$ for two MWNTs [20]. For the self-retracting motion to take place, the retraction driving force $F_d^*$ must exceed the resistance force, $F_r$. The resistance force is equal to $L(L - x)\tau_s$ for initiating the retracting motion or $L(L - x)\tau_d$ during the retracting motion, where $\tau_s$ and $\tau_d$ denote the static and dynamic graphite shear strengths between non-rotated graphite layers. The values of $\tau_s$ were measured to arrange from 0.27 to 0.75 MPa (mean value 0.48 MPa [21]) depending on sliding direction, and the values of $\tau_d$ were found to be one order in magnitude smaller than those of $\tau_s$. Using the initiation condition of the self-retracting motion: $F_d^* > F_r$ or $L - x < 2\gamma^*/\tau_s$ and taking the intermediate value $\gamma_e = 0.56$ Jm$^{-2}$, we have obtained the estimate that the critical side length $L_{cr} = 2\gamma^*/\tau_s$ ranges from 0.84 to 2.1 μm for non-rotated microflakes. This agrees well with our observations. We further examine the self-retracting condition for a rotated microflake. We should expect that the retraction driving force for a rotated microflake is a little bit smaller than but in the same order as that for a non-rotated one. However, the static graphite shear strengths $\tau_s^*$ between rotated graphite layers are one order smaller than $\tau_s$. These imply that the self-retracting motion may occur for rotated microflakes with side lengths of several to dozens micrometers.

To understand why in our experiments the self-retracting motion was observed only for few 5 μm microflakes, we lifted a 5 μm non-self-retracted microflake and noted its warped shape [Fig. 3(b)]. This indicates that the microflake had experienced a deformation exceeded the elastic range. Because the retraction force, $F_d$, results from the van der Waals interaction, it reduces rapidly to zero as the separation spacing between the microflake and the platform increases from the optimized value $s_0 = 0.335$ nm. For example, using the 6-12 Lennard-Jones potential we estimated $F_d$ at $s = 2$ nm to be less than 1% of the value of $F_d$ at $s = s_0$. Similarly, the edge-effect force reduces rapidly with the increasing separation spacing. Consequently, the total retraction driving force $F_d^*$ of a warped microflake would be much smaller than that of a flat microflake. Thus, it is likely that $F_d^*$ of the observed warped microflakes in our experiment were too small to drive the retracting motion. On the other hand, to slip out a microflake, the applied force through the micromanipulator probe must exceed the sum of the static interlayer shear strength force, $F_r = L^2\tau_s$, and the edge interlayer dangling bond interaction, $F_{db} = 4L\tau_{db}$, where $\tau_{db}$ characterizes the edge interlayer dangling bond interaction strength. The initiation of slipping is thus most likely to occur between two adjacent layers that have the smallest $\tau_{db}$.

Because the resistance force, $L^2\tau_s + 4L\tau_{db}$, against initiating slip of microflakes increases with the increasing microflake size $L$, and because the thickness of the $SiO_2$ coats -- the main bodies against warping deformation, and their elastic limits are fixed, the permanent deformations, like warping, of microflakes slipped in our experiments should be size-dependent and occurred only for larger microflakes. Consequently, slipped microflakes with size larger than 5 μm would be more severely distorted from the flat configuration, and they thus are much unlikely to self-retracting. This explanation does not contradict to the prediction of our above analysis that larger microflakes, up to several dozens of micrometers in size, may still experience self-retracting motion if they would remain flat and were rotated.

What we have presented above is the first observation, to the best of our knowledge, of the self-retracting motion of graphite microflakes slipped from graphite platforms. We expect that the self-retracting motion of the same nature occur in other lamellar solids that are of super-low inter-lamellar shear resistance strengths, such as molybdenum disulphide, Biotite, and Phlogopite. Furthermore, we expect to observe the self-retraction motion if one places a carbon nanotube on a graphite platform and pushes it to a suspended position. The findings reported in this Letter may conduce to create nano-electromechanical systems with a wide range of mechanical operating frequency from mega to giga hertzs. In fact, with the same physical principle for predicting the ultra fast MWNT oscillator [15], the self-retracting motion of graphite microflakes could be used to fabricate oscillators with frequencies in a wider range than that of MWNT oscillators. For instance, the oscillation frequency $f$ of an $N$-layer graphite microflake with the oscillation amplitude $\Delta$ on the square graphite platform of side length $L$ is estimated to be about $f = 100/\sqrt{NL\Delta}$. It ranges from 5.8 MHz as $L$ = 3 μm, $\varDelta = L/3$, and $N = 100$ (thickness = 33.5 nm) to 1.0 GHz as $L = 100$ nm, $\varDelta = L/3$, and $N = 3$ (thickness = 1 nm).

The work is supported by the National Science Foundation of China (NSFC) through Grant Nos. 0332020, 10121202, and 10672089, and NSFC/RGC through Grant No. 50518003.

**Supporting Online Material**

**Movie S1:**

A forced-sliding and retracting motion of a graphite microplate operated by using the probe tip of a manipulator within a SEM. (MPEG; 3.2MB)

**Movie S2:**

The self-retracting motion of a slid graphite microplate upon vertically withdrawing the probe. (MPEG; 2.06MB)